# Impact of CIR Storms on Thermosphere Density Variability during the Solar Minimum of 2008

Jiuhou Lei • Jeffrey P. Thayer • Wenbin Wang • Robert L. McPherron

―――――――

J. Lei • J.P. Thayer

Department of Aerospace Engineering Sciences, University of Colorado, Boulder, CO, USA
Email: Jiuhou.Lei@Colorado.edu

W. Wang

High Altitude Observatory, National Center for Atmospheric Research, Boulder, CO, USA

R.L. McPherron

University of California Los Angeles, Institute Geophysics and Planetary Physics, Los Angeles, CA, USA

**Abstract.** The solar minimum of 2008 was exceptionally quiet, with sunspot numbers at their lowest in 75 years. During this unique solar minimum epoch, however, solar wind high‒speed streams emanating from near-equatorial coronal holes occurred frequently and were the primary contributor to the recurrent geomagnetic activity at Earth. These conditions enabled the isolation of forcing by geomagnetic activity on the preconditioned solar minimum state of the upper atmosphere caused by Corotating Interaction Regions (CIRs). Thermosphere density observations around 400 km from the CHAMP satellite are used to study the thermosphere density response to solar wind high‒speed streams/CIRs. Superposed epoch results show that thermosphere density responds to high‒speed streams globally, and the density at 400 km changes by 75% on average. The relative changes of neutral density are comparable at different latitudes, although its variability is largest at high latitudes. In addition, the response of thermosphere density to high‒speed streams is larger at night than in daytime, indicating the preconditioning effect of the thermosphere response to storms. Finally, the thermosphere density variations at the periods of 9 and 13.5 days associated with CIRs are linked to the spatial distribution of low‒middle latitude coronal holes on the basis of the EUVI observations from the STEREO.

# 1. Introduction

The solar minimum of 2008 was exceptionally quiet, with sunspot numbers at their lowest since 1913. This placed the ionosphere and thermosphere properties in a steady, low solar flux preconditioned state. It was expected that this deep solar minimum would provide a unique opportunity to minimize the magnetospheric and solar-driven disturbances to study dynamics in the ionosphere – thermosphere system driven from the lower atmosphere (Sojka *et al.*, 2009). However, accompanying this epoch of low, nearly constant solar EUV flux were recurrent high – speed solar wind streams and the resultant geomagnetic disturbances associated with large, long-lived low latitude solar coronal holes (Gibson *et al.*, 2009).

Coronal holes of substantial size near the Sun – Earth line can create high – speed solar wind. High speed solar wind can interact with the slower solar wind in interplanetary space leading to disturbances in the solar wind called corotating interaction regions (CIRs) (*e.g.*, Tsurutani *et al.*, 1995, 2006). These disturbances lead to modest enhancements in Earth's geomagnetic activity with typical Kp values around four. Additionally, coronal holes can persist for many solar rotations in the solar atmosphere. Consequently, solar wind disturbances occur at harmonics of the solar rotation period leading to recurrent geomagnetic activity. The recurring periodicity in CIRs enables a correlation analysis with periodicities in geomagnetic activity and changes in the ionosphere – thermosphere system. Oscillations at multi-day periods (near 7 and 9 day subharmonics of solar rotation) associated with recurrent high – speed solar wind streams were recently discovered in the thermosphere and ionosphere properties during the declining phase of Solar Cycle 23 (Lei *et al.*, 2008a, 2008b, 2008c; Mlynczak *et al.*, 2008; Thayer *et al.*, 2008; Crowley *et al.*, 2008).

One important outcome of Lei *et al.* (2008a, 2008c) and Thayer *et al.* (2008) was that the solar EUV flux did not vary appreciably at these shorter sub-harmonic periods suggesting that geomagnetic forcing on the thermosphere could possibly be isolated. However, this discovery was found during the declining phase in the solar cycle such that the solar EUV flux was gradually decreasing over multiple periods of recurrent geomagnetic activity, thus changing the base state of the thermosphere over time and leading to a different response for each particular storm. The solar minimum period in 2008 showed nearly constant EUV flux over many recurrent geomagnetic storms enabling isolation and identification of the thermospheric response solely to CIR-forced geomagnetic activity. In this study, we will investigate the thermosphere response in the solar minimum 2008 to CIR storms, including its latitude and local time dependence in a quantitative way through a superposed – epoch analysis.

## 2. Data Sets

The primary data source used in this work is thermospheric density from the *Challenging Minisatellite Payload* (CHAMP) satellite. The *Spatial Triaxial Accelerometer for Research* (STAR) Level 2 observations were processed by the GeoForschungsZentrum Potsdam (GFZ) to remove maneuvers and anomalous spikes and smoothed over 10s. Thermosphere mass densities from pole to pole are then obtained from accelerometer measurements using standard methods (Sutton, Nerem, and Forbes, 2007). The measured densities at satellite altitudes are normalized to a constant altitude of 400 km using NRLMSISE00 (Picone *et al.*, 2002), and this normalization process is necessary in order to minimize the effect of slight changes of the satellite altitude.

In addition, the solar wind and Interplanetary Magnetic Field (IMF) parameters obtained from the OMNI database (http://omniweb.gsfc.nasa.gov/) are used to identify stream interfaces within CIRs in 2008. The interface is defined by a variety of characteristics including compression and flow deflection. See details in McPherron, Baker, and Crooker (2009). The time of the zero crossing of the azimuthal flow angle is taken as that of the interface, which represents the CIR zero epoch time in the next section.

## 3. Results

Figure 1 from top to bottom shows solar wind density $N_n$, interplanetary magnetic field magnitude B, solar wind speed, interplanetary magnetic field $B_z$ and $B_x$ components in GSM coordinates, geomagnetic activity index Kp, and CHAMP ascending orbit-averaged neutral density at 400 km in 2008. The most striking feature in solar wind density $N_n$ is the sharp spikes, closely followed by the peaks in B. The subsequent enhancement in solar wind speed can be seen clearly, which follows each peak in $N_n$ and B. In addition, $B_z$ shows rapid fluctuations with a typical amplitude of ~8 nT in the stream–stream compression region and around 2–4 nT in the high-speed streams in a higher time–resolution plot (not shown). These are the classical features of a CIR as described by Tsurutani *et al.* (1995, 2006; and references therein). There were 38 high–speed streams/CIRs in the entire year of 2008, according to the criterion of McPherron, Baker, and Crooker (2009). Additionally, $B_x$ oscillates in direction alternately between "toward" (+) and "away" (-) from the Sun in each Carrington Rotation (CR) period delimited by vertical lines, which is suggestive of the dominant two-sector structure. The corresponding oscillations with multi-day periods are clearly seen in Kp as a result of the CIR storms. Consequently, thermosphere densities from CHAMP present high-frequency peaks (within each solar–rotation period), which are imbedded in the larger time–scale variations due to the seasonal variation of the thermosphere and local time progression of the satellite.

The Lomb – Scargle (Lomb, 1976; Scargle, 1982) amplitude spectra of the corresponding parameters in Figure 1 are presented in Figure 2 including thermosphere density from CHAMP descending-orbits as well. Obviously, all variables display strong 9- and 13.5-day periodicities except that the 9-day periodicity is absent in the IMF $B_x$ and $B_z$ components. The 9-day periodicity is also absent in $B_y$ (no plotted), but it is strongly present in the total B (Figure 2b) due to compression effects at the beginning of the CIRs (see Figure 1). Note that the 9- and 13.5-day periodicities are not due to mathematical artifacts since the oscillations at these periods are clearly seen in the time series (Figure 1). The results of this study indicate again that these strong 9- and 13.5-day oscillations in the thermosphere are associated with the solar-wind energy input, which is in accordance with our previous studies (*e.g.*, Lei *et al.,* 2008a, 2008c; Thayer *et al.*, 2008).

Figure 3 shows the same variables as those in Figure 1, except zooming into the two solar rotation periods of CR2068 and CR2075 in Spring and Fall seasons, respectively. Again, the CIR storms in both CR2068 and CR2075 share the common features with those in the schematic diagram of Tsurutani *et al.* (1995, 2006). It should be pointed out that CR2068 is the focus period of the Whole Heliosphere Interval (WHI) campaign (Gibson *et al.*, 2009). Three high–speed streams occurred during the WHI period. The oscillations with the dominant periods of 9 and 13.5 days also prevailed in solar wind, geomagnetic activity, and neutral density in this interval. Furthermore, they were recurrent at least for the first half of 2008 as seen in Figure 1. In CR2075, four streams, which centered on the day of year (DoY) around 275, 285, 293, and 300, were observed and the second stream only lasted around three days within the individual solar rotation in comparison with lasting around 13 days in CR2068. In addition, in CR2068 the first stream is more geoeffective as indicated by Kp and neutral density response (Figure 3f–g) when $B_x$ has the "toward" (+) sector (Figure 3e) around the Spring equinox and $B_z$ tends to have larger southward component (Figure 3d). In contrast, the second stream, when $B_x$ has the "away" (-) sector (Figure 3e'), is more geoeffective (Figure 3f'–g') in CR2075 around the fall. This indicates that the Russell–McPherron effect (Russell and McPherron, 1973) may play a role in determining the geoeffectiveness of a high–speed stream/CIR.

To examine the thermosphere response to the CIRs quantitatively, we carried out a superposed epoch analysis on 29 of 38 total CIRs that were isolated events with a duration of more than three days, regardless of recurrence of the stream interfaces. The percentage changes of both ascending and descending orbit-averaged neutral densities (equivalent to 58 events for thermosphere density) during the CIRs to the reference of one day prior to the CIR interfaces are computed to minimize the seasonal and local time effects. In Figure 4, from top to bottom, are the superposed epoch results for solar wind density $N_n$, solar wind speed, solar wind dynamic pressure, $B_z$ component, Kp, and the relative changes in neutral density for both ascending and descending orbits. The initial, main, and recovery phases of the CIR storm defined in Tsurutani *et al.* (1995) can be seen from this figure. Both solar wind density $N_n$ and dynamic pressure show a strong increase during the initial phase of the CIR associated with the crossing of the heliosphere current sheet. $B_z$ has a weak southward component (in an averaged sense) at the stream interface and it shows larger variability during the main phase because of the stream-stream compressive effects. Kp increases rapidly during the initial phase. However, it is unclear why the increase of neutral density during the initial phase is not as sharp as that seen in Kp. This is probably because the geomagnetic activity may start in the midnight sector auroral zone and expand to the dayside. Neutral density takes about 7–8 days to recover back to its value under quiet conditions, which is the imprint of the long recovery of the solar wind speed and the resultant geomagnetic activity. Interestingly, the relative change of neutral density to the CIR is as large as 75% on average, which indicates that high–speed streams/CIRs have a significant impact on the day-to-day variability of the thermosphere and satellite drag even during the "quiet Sun" period, although the resultant geomagnetic activity is weak or moderate.

The CHAMP density measurements from pole to pole also allow us to further investigate the latitudinal dependence of the thermosphere response to CIRs. Although its variability (*i.e.*, the spread of the standard deviation) is largest at high latitudes (Figures 5a–c), it is surprising that the relative changes of neutral density are comparable at different latitudes (Figures 5d–e). In addition, the nighttime density at 400 km (Figure 5e) increases by 85% on average during the main phase, which is larger than the increase in daytime, indicating the preconditioning effect of the thermosphere response to storms, as explained next. Lei *et al.* (2010) showed that the increase of neutral temperature during geomagnetic storms results in the integration effects of the changes of all scale heights below the satellite altitude. Consequently, the increase in neutral density is larger at night when the scale height of the neutral gas is smaller than that in daytime because neutral density falls off rapidly as a function of scale height. It is clear in Figure 5 that neutral density is quietest one day prior to the stream interface, which accords well with the geomagnetic calm before the CIR storm (Figures 1 and 4) (also see Tsurutani *et al.*, 1995, and Borovsky and Steinberg, 2006). An important application of this result is that the ionosphere-thermosphere observations one day prior to the stream interface might be more suitable to minimize the effect of the magnetospheric energy input to investigate the variability of the ionosphere – thermosphere solely due to the influence from the lower atmosphere or anthropologic sources.

## 4. Discussion

Recent studies demonstrated that the 9-day solar wind variations in 2005, which contribute to the corresponding oscillations in the thermosphere and ionosphere (Lei *et al.*, 2008a, 2008b; Mlynczak *et al.*, 2008), are due to the existence of a triad of solar coronal holes distributed roughly 120 degrees apart in solar longitude (Temmer, Vršnak, and Veronig, 2007). The remaining question is what spatial distribution of coronal holes results in the strong 9- and 13.5-day periodicities in solar wind parameters and thermosphere density in 2008.

Figure 6 illustrates the distribution of coronal holes made from the STEREO-A EUVI 195Å measurements as a function of solar latitude and Carrington longitude along with temporal variations of solar wind speed and thermosphere density during the WHI period. There were two large equatorial – low latitude coronal holes marked by the red circles. The coronal hole in the northern hemisphere is responsible for the high-speed stream during DoY 86 – 95 (around the period of 9 days) and the broader coronal hole in the southern hemisphere is the cause of the high–speed stream during DoY 95 – 107 (near the period of 13.5 days). It is immediately apparent that the 9- and 13.5-day periodicities in solar wind and thermosphere density during WHI (Figures 6b – d) are associated with these two coronal holes extending from the northern and southern hemispheres, respectively. In addition, the third stream during DoY 107 – 113 is associated with the coronal hole from higher latitudes in the southern hemisphere. Thereby, both the width and separation of low latitude coronal holes play an important role in determining the temporal variations of high–speed streams, geomagnetic activity, and the corresponding geospace response.

Finally, it is worth noting that the coronal holes shown in Figure 6a actually lasted over several solar rotations as seen in the STEREO EUVI measurements (http://secchi.nrl.navy.mil/synomaps), and these long-lasting coronal holes contributed to the high–speed streams/CIRs which occurred frequently in the deep solar minimum of 2008. Additionally, the low–middle latitude coronal hole from the southern hemisphere was diminished in the second half of 2008. Therefore, the resultant high–speed streams evolved as peaks with much shorter duration in comparison with the corresponding streams in the first half of 2008 (see Figure 1).

## 5. Conclusion

During the unique solar minimum of 2008 when solar EUV flux was low and nearly constant, solar wind high–speed streams emanating from near-equatorial coronal holes occurred frequently and were the primary contributor to the continuous geomagnetic activity at the Earth. These conditions enabled the isolation of forcing by geomagnetic activity on the preconditioned solar minimum state of the upper atmosphere caused by CIRs. Intense magnetic field regions or CIRs are due to the interaction of high–speed corotating streams with the upstream low speed solar wind, and high density heliospheric current sheet. There were 38 high–speed streams/CIRs in 2008. Thermosphere density observed from CHAMP was significantly disturbed due to CIRs even though their resultant recurrent geomagnetic activity was weak or moderate. Therefore, the variations of thermosphere density provide a potential trace of solar wind/magnetospheric energy input into the upper atmosphere.

The main conclusions of this study are summarized as follows: ⅰ) Superposed epoch results have shown that thermosphere density responds to high–speed streams globally and the density at 400 km changes by 75% on average. ⅱ) The relative changes of thermosphere density are comparable at different latitudes, although its variability is larger at high latitudes. ⅲ) Thermosphere density has a larger response (in a relative sense) at night than in daytime due to their differing scale heights. ⅳ) The thermosphere is calmest one day prior to the stream interface. ⅴ) According to the STEREO EUVI observations, the thermosphere density variations at the periods of 9 and 13.5 days associated with CIRs are linked to the spatial distribution of low-middle latitude coronal holes.

**Acknowledgments**. This work was in part supported by NASA NNX10AE62G, AFOSR MURI Award FA9550-07-1-0565, and NASA LWS NNX08AQ91G. One of the authors (RLM) would like to acknowledge support provided by a NASA Grant NNX20AE61G, and additional support was provided at LASP in Boulder, CO by the CISM project through an STC Program of the National Science Foundation under Agreement no. ATM-0120950. We thank Eric Sutton for his assistance with CHAMP data, and we also wish to thank Barbara Emery and Editor John Leibacher for their comments on the draft. The STEREO EUVI data are provided by the STEREO science center http://stereo-ssc.nascom.nasa.gov/.

# Figure Caption

**Figure 1**. Variations of solar wind density $N_n$, interplanetary magnetic field B, solar wind speed, interplanetary magnetic field $B_z$ and $B_x$ components in GSM, geomagnetic activity index Kp, and CHAMP ascending orbit-averaged neutral density at 400 km in 2008. Note that ascending and descending orbits of the CHAMP satellite are distinguished because they have different local time sampling and thermosphere density strongly depends on local time. The vertical dashed lines indicate the start times of the Carrington Rotations. CR2068 and CR2075 are marked in this figure. Four days have been added to the solar rotation start times to account for the travel time from Sun to Earth.

**Figure 2**. Lomb–Scargle spectra of solar wind density $N_n$, interplanetary magnetic field B, solar wind speed, interplanetary magnetic field $B_z$ and $B_x$ components in GSM, geomagnetic activity index Kp and CHAMP ascending and descending orbit-averaged neutral density at 400 km.

**Figure 3**. Same as Figure 1, but for CR2068 (left) and CR2075 (right) in Spring and Fall seasons, respectively.

**Figure 4**. Superposed epoch results of CIRs for solar wind density $N_n$, solar wind speed, solar wind dynamic pressure, $B_z$ component, and Kp along with the relative changes in neutral density. The heavy solid lines define the median, and the shaded area represents the standard deviation. See details in the text.

**Figure 5**. Superposed epoch results of the relative changes in neutral density at high (60–90 degree, a), middle (30–60 degree, b) and low (0–30 degree, c) latitude bands, the averaged changes of density for the three latitude bands (d), and also the density response to CIRs (e) in daytime (dotted lines) and at night (solid lines). The heavy solid lines in panels a–c define the median, and the shaded area represents the standard deviation.

**Figure 6**. Spatial distribution of coronal holes from the STEREO EUVI 195Å measurements as a function of solar latitude and Carrington longitude along with temporal variations of solar wind speed and thermosphere density in CR2068. Note that the STEREO EUVI A was "ahead" of the Earth during this period (~2 days away from the Sun–Earth line). The time runs from right to left for the time series of solar wind speed and thermosphere density. Four days have been added to the solar rotation start times of CR2068 to account for the travel time from Sun to Earth.

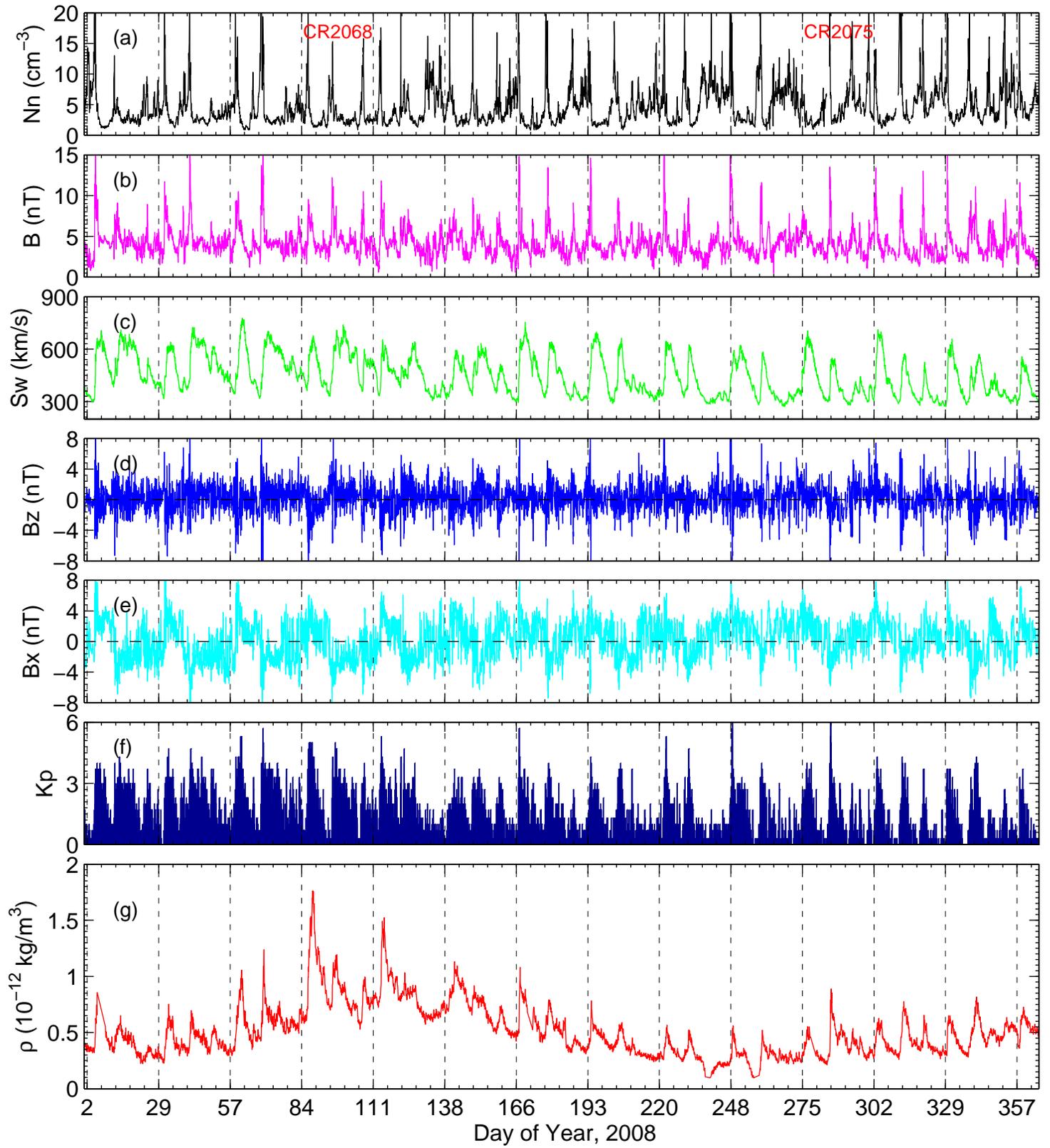

Figure 1

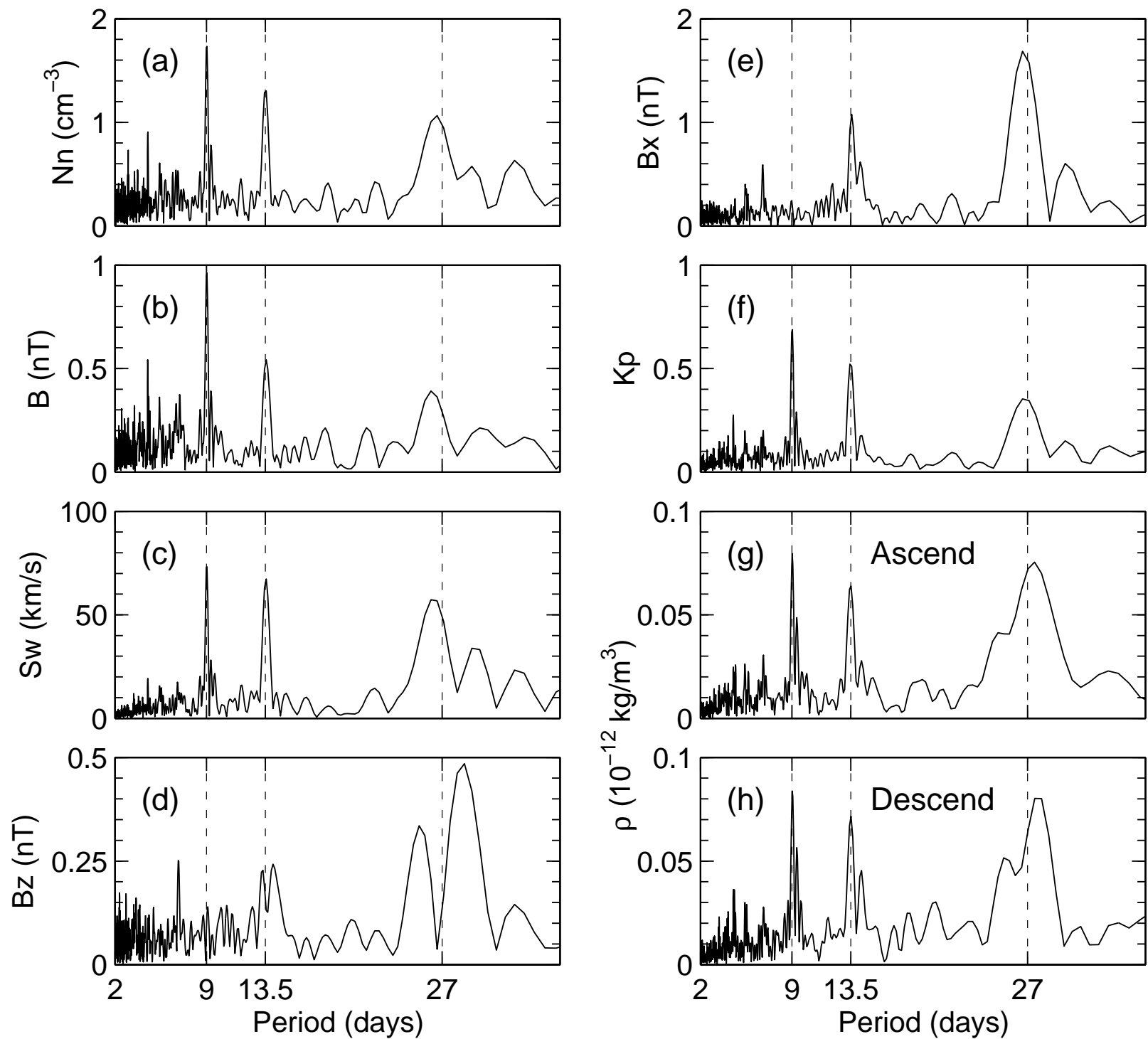

Figure 3

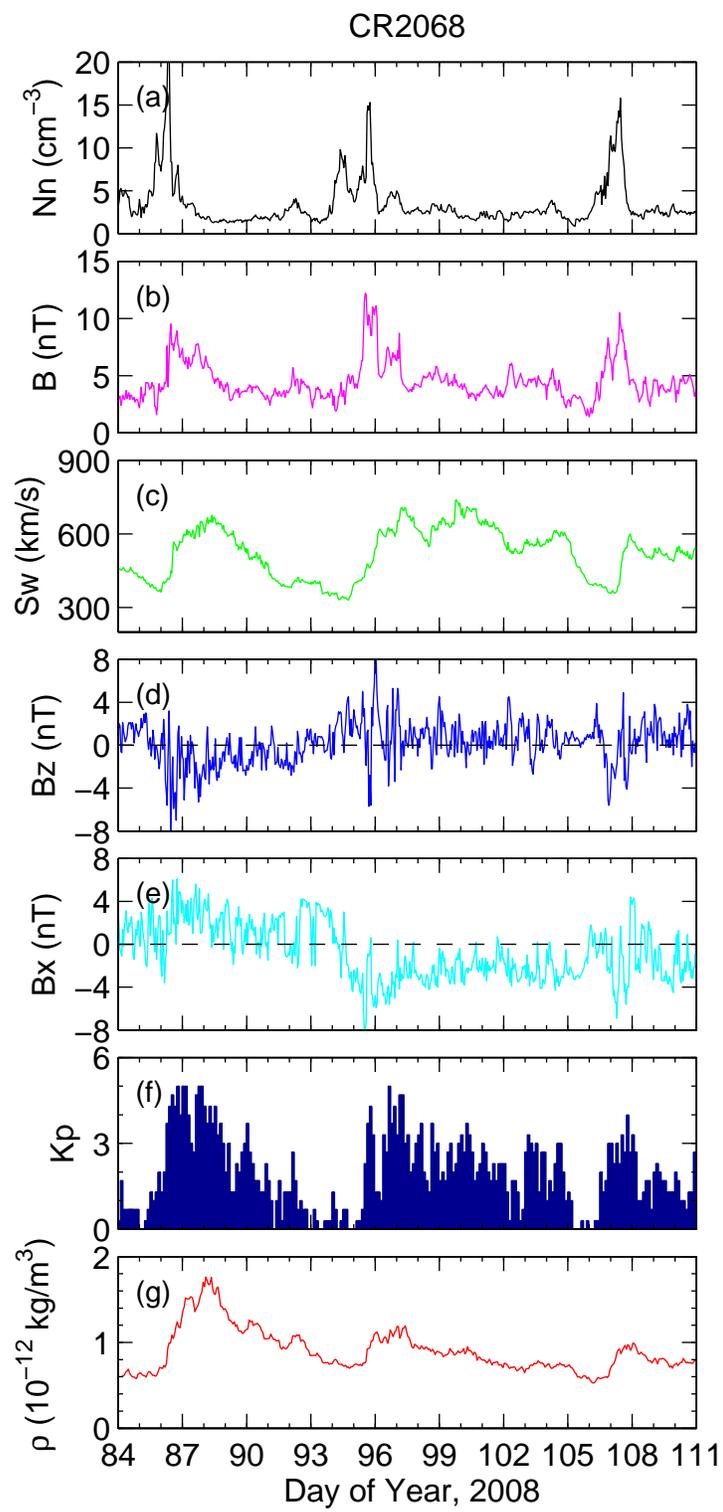
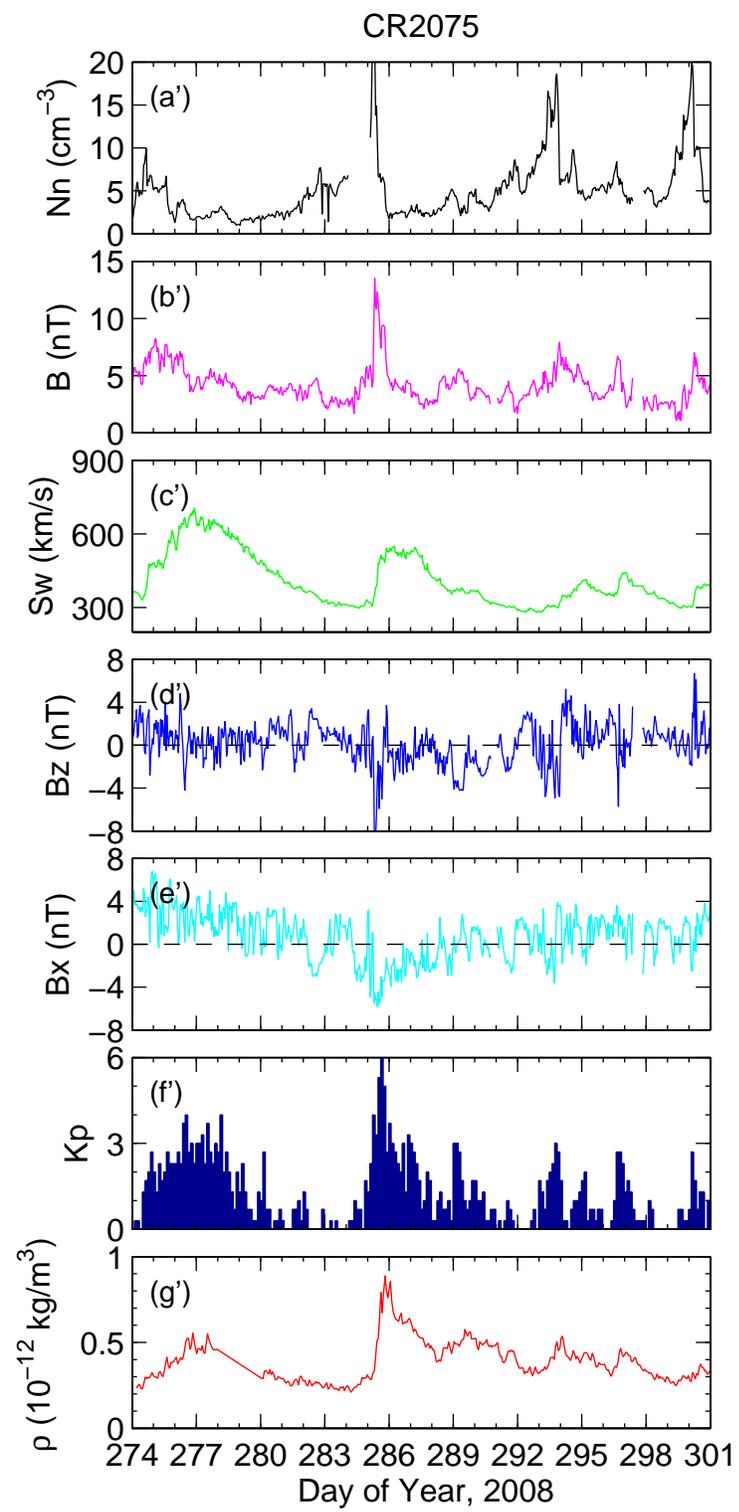

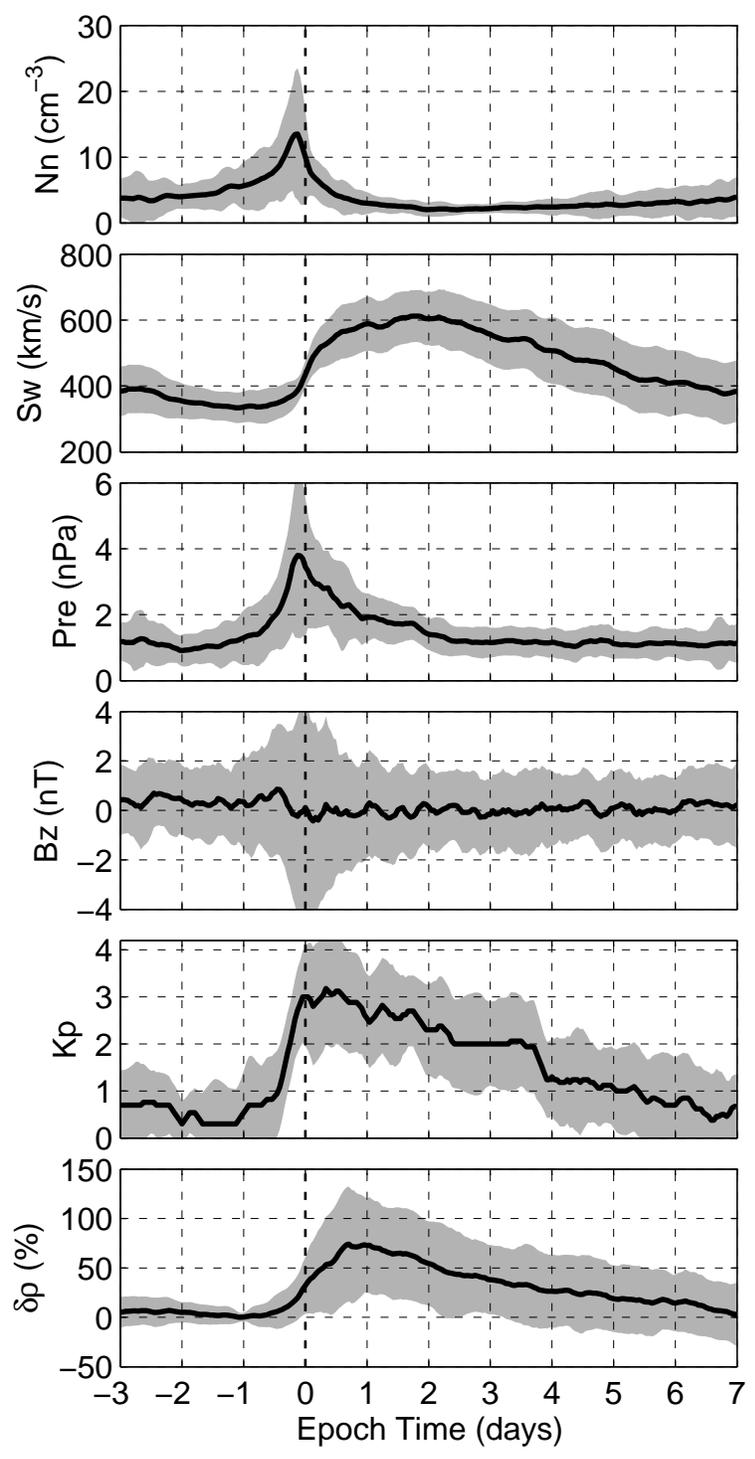

Figure 5

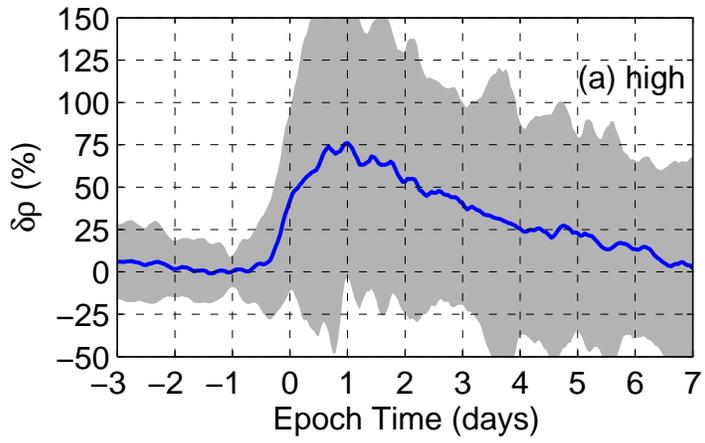
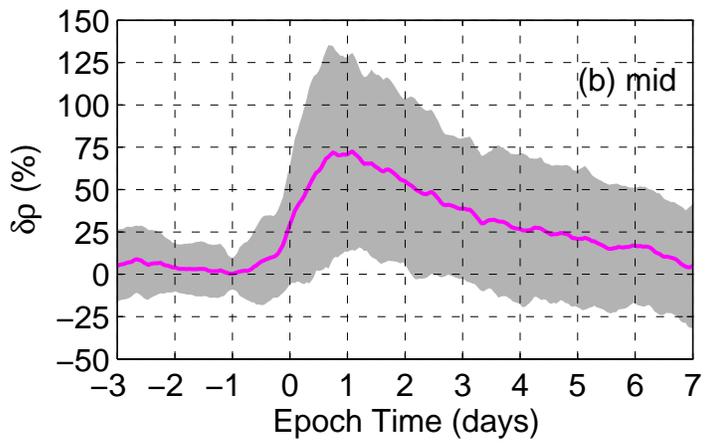
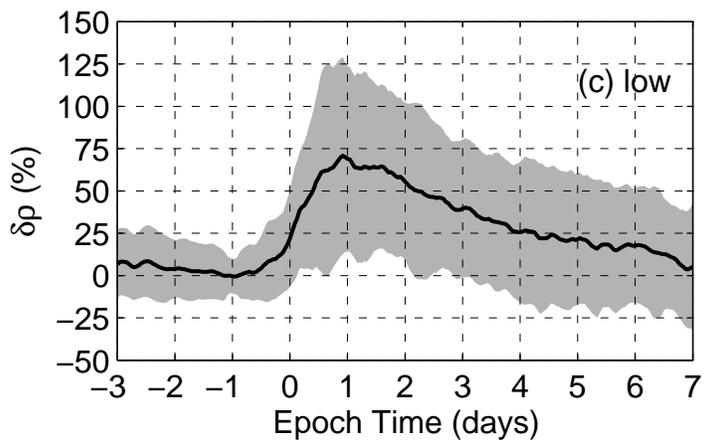
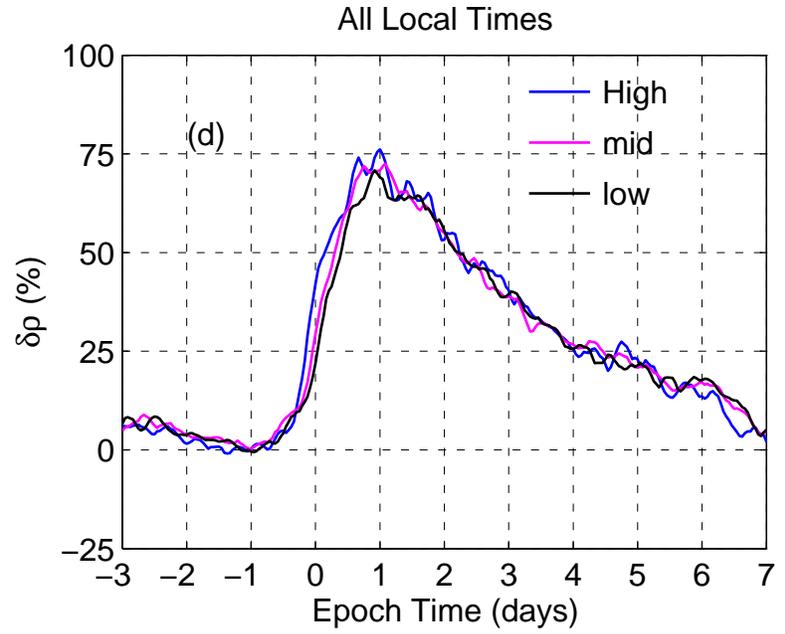
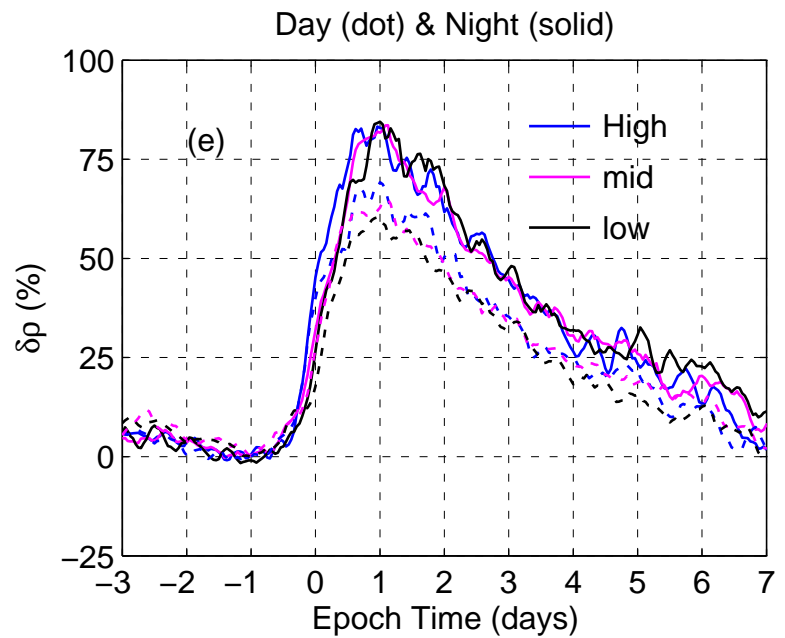

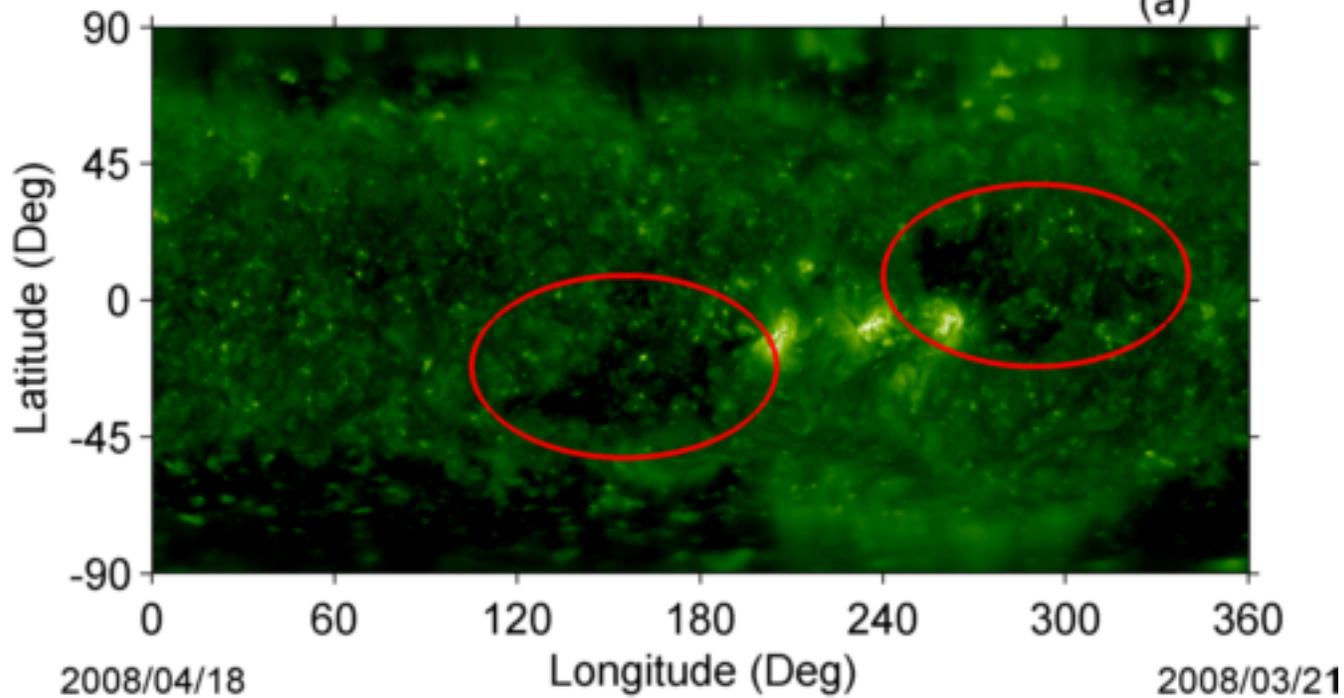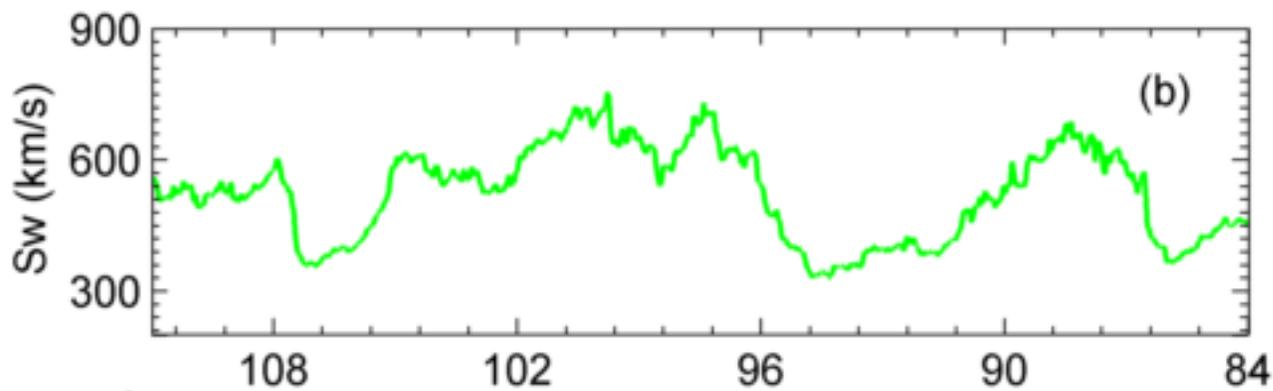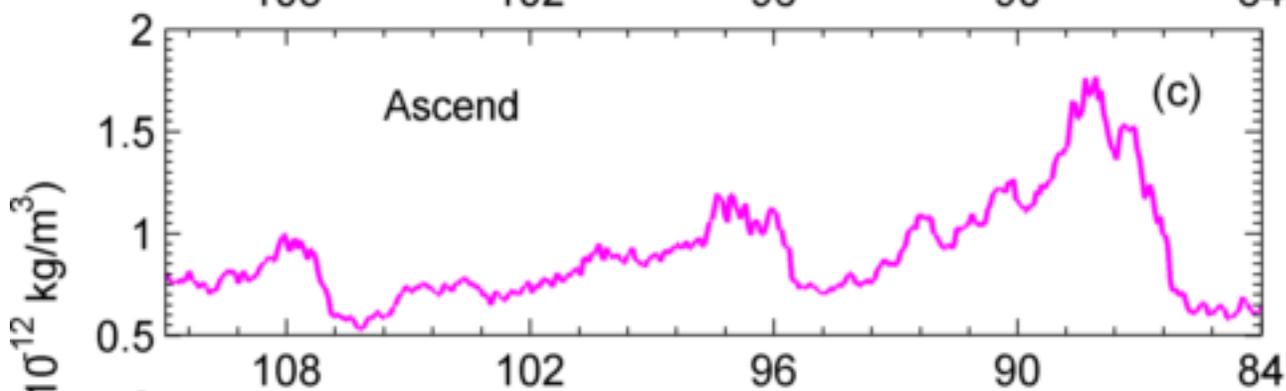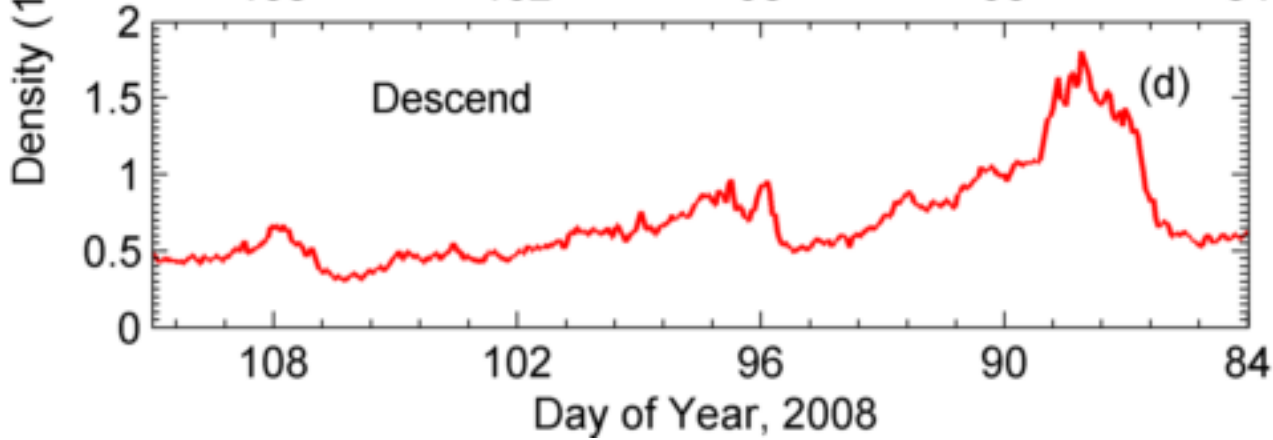